\documentclass[conference]{IEEEtran}
\IEEEoverridecommandlockouts

\usepackage{cite}
\usepackage{amsmath,amssymb,amsfonts}

\usepackage{graphicx}
\usepackage{textcomp}
\usepackage{xcolor}
\usepackage[ruled,vlined]{algorithm2e}
\usepackage{algpseudocode}
\usepackage[para]{footmisc}
\usepackage{tabularray}
\def\BibTeX{{\rm B\kern-.05em{\sc i\kern-.025em b}\kern-.08em
    T\kern-.1667em\lower.7ex\hbox{E}\kern-.125emX}}
\begin{document}

\title{Orchestrating the Execution of Serverless Functions in Hybrid Clouds}

\author{
\IEEEauthorblockN{Aristotelis Peri, Michail Tsenos, Vana Kalogeraki}
\IEEEauthorblockA{Athens University of Economics and Business\\ Athens, Greece\\
\{ariperi, tsemike, vana\}@aueb.gr}
}

\maketitle

\begin{abstract}

In recent years, serverless computing, especially Function as a Service (FaaS), is rapidly growing in popularity as a cloud programming model. The serverless computing model provides an intuitive interface for developing cloud-based applications, where the development and deployment of scalable microservices has become easier and cost-effective. An increasing number of batch-processing applications are deployed as pipelines that comprise a sequence of functions that must meet their deadline targets to be practical. 
In this paper, we present our Hybrid Cloud Scheduler (HCS) for orchestrating the execution of serverless batch-processing pipelines deployed over heterogeneous infrastructures. Our framework enables developers to (i) automatically schedule and execute batch-processing applications in heterogeneous environments such as the private edge and public cloud serverless infrastructures, (ii) benefit from cost reduction through the utilization of their own resources in a private cluster, and (iii) significantly improves the probability of meeting the deadline requirements of their applications. 
Our experimental evaluation demonstrates the efficiency and benefits of our approach.

\end{abstract}


\section{Introduction}

One of the newest and most enticing Cloud computing models in recent years is the Serverless computing paradigm. 
This new paradigm shifts the complexity of allocating and provisioning resources to the Serverless provider, which aims to provide the illusion of always-available resources. In Serverless environments, developers write applications as Functions. Each Function and its dependencies are packed and delivered typically as a container and run in a Serverless platform. Well-known commercial Serverless platforms include AWS Lambda\footnote{https://aws.amazon.com/lambda/} and Google Cloud Functions\footnote{https://cloud.google.com/functions}. 
These platforms take over tasks like provisioning, managing and scaling where users only need to provide code, in any programming language, and run it, typically in response to event triggers.
Serverless Platforms such as Google Cloud Run\footnote{https://cloud.google.com/run} allow the developers to upload their own Docker Containers and also allow more flexibility to the developer. At the same time, open-source Serverless platforms such as OpenFaas\footnote{ https://www.openfaas.com/} allow organizations to use their private infrastructures for hosting Serverless applications. 
The Serverless computing model has found popular applications in a wide variety of settings, from data analytics to stream processing at scale, multimedia delivery and developing chatbots.
Edge computing offers companies the opportunity to create
small private edge clusters at the edge of the network
to facilitate the needs of their employees
or act as offloading units for low-latency demands, 
thus reducing the bandwidth load of the network core.
Edge computing complements the Cloud platforms by pervasively deploying compute nodes placed between the central Cloud infrastructure and the end devices 
({\it e.g.,} IoT devices) to offer low-latency and secure services.
This new hybrid cloud infrastructure supports pushing computation towards a continuum from the cloud to the edge, where the user devices are located, as opposed to primarily taking place in centralized cloud environments. 

The newly hybrid cloud offers possibilities for a runtime system that can utilize to its advantage the plethora of a) different computing and storage infrastructures, b) data locations and c) processing frameworks that can be adaptively utilized and seamlessly combined to perform efficient computations at scale while ensuring seamless deployment.
Even though the hybrid cloud is a promising approach for latency reduction, there are several challenges that still need to be addressed associated with the deployment over heterogeneous resources and meeting application latency requirements, as (a) real-world serverless functions are characterized by skewed invocations, burstiness and highly variable execution times, (b) they are built over dedicated clusters and resources where users utilize only a few nearby edge options that can provide a low-latency response, and (c) overloads can easily happen since dedicated resources are physically limited and lack scaling capabilities.

In this paper, we propose a Hybrid Cloud runtime system for orchestrating the execution of serverless functions across public and private infrastructure. Our focus is on providing an efficient, elastic, and cost-effective execution environment, along with distributed scheduling techniques for containerized tasks dispatched over different platforms and hardware infrastructures. This allows us to leverage the full potential of Hybrid Cloud and reduce costs by utilizing private edge clusters effectively.
To achieve this, we have developed a two-level scheduling approach called the Hybrid Cloud Scheduler (HCS scheduler). In the first level, we determine the scheduling policy, and in the second level, we make placement decisions in the edge cluster to meet application requirements and maximize system utilization. The HCS scheduler works in conjunction with a Serverless Batch Processing Engine, which allows users to schedule and execute batch jobs automatically without knowledge of the underlying environment or allocated resources. Users only need to provide Docker images containing the code for batch-processing steps and design a Directed Acyclic Graph (DAG) representing the workload pipeline, along with setting the desired job execution deadline. The HCS Scheduler automatically performs the scheduling of the DAG.
Overall, our approach offers an efficient and flexible solution for executing serverless functions in Hybrid Clouds, leveraging different execution environments and maximizing resource utilization in private edge clusters.

In this paper we make the following contributions: 
\begin{itemize}

\item We propose a unified Hybrid Cloud system comprising a plethora of computing and storage resources that are utilized to support the execution of Serverless functions. 

\item We take a systematic approach and create a scheduling policy derived from recent works on cluster, task and Serverless scheduling frameworks that aim to jointly satisfy application requirements (e.g., reduce latency) and system requirements (e.g., maximize resource utilization).

\item We have designed and implemented a Driver program which enables the user to express a workload pipeline as a DAG of Serverless functions by providing the batch-processing steps as Functions packaged within Docker images, the minimum required replicas of each step and a desired execution deadline and then execute it at scale.
\item We provide a rich evaluation of our proposed scheduling policy for scheduling functions in Hybrid Clouds, considering the application characteristics, resource availability and the trade-offs involved. Our experimental results illustrate that our hybrid execution environment for batch-processing offers up to 87\% cost reduction compared to executing the same workloads in the public Cloud while meeting a very high percentage of SLOs.

\end{itemize}

\section{System Model}

\subsection{Data Model}
\label{dataModel}

Let a set of batch jobs where each batch job processes a set of data being stored in a storage system such as S3. We consider that the data is fragmented in blocks of fixed size $S$ bytes. Our processing model, similar to the popular Serverless systems available in the Cloud, respects the following two parameters: (i) The processing of each fragment should start with an HTTP call to the function, and (ii) the duration of the processing of each fragment should not exceed the processing time limit defined by each system. 
Thus, the size $S$ of each fragment depends on this time interval. In the case of very large fragments, it is possible that the execution time of some requests exceeds the predefined execution timeout (e.g 60 seconds) of the system, in which case the requests will be rejected.
During execution, the function receives via HTTP a JSON object that describes the location of the fragment (S3 Bucket Id and Object Id). The function gets the fragment from this location and performs the predefined processing and then stores the result back in storage. Finally, it returns the location of the result as a response to the HTTP request. In the case that it does not produce results, it uses the X-NoOp Header.

\subsection{Execution Model}
We consider that the system receives requests for processing
a set of Batch Jobs. Each Batch Job (BJ) must be completed within a predefined time given by the user. This time period is a soft deadline, if it is violated this is not catastrophic for the system.
In the Serverless execution model, the data is processed via a Pipeline. A Pipeline denotes a sequence of containerized Serverless function invocations, typically triggered by a Message. The Pipeline is represented as a Directed Acyclic Graph (DAG), where nodes correspond to Serverless functions, named Processing Steps (PS), and the edges in the DAG denote the corresponding invocations. Each Batch Job has its own unique Id.

Each Serverless Function uses the Data Model described in the previous Section \ref{dataModel}. 
Each \emph{PS} can serve one function only and has a unique Id.
At each \emph{PS}, the system can instantiate a number of active function replicas and speed up the processing of the step by distributing the workload to the replicas.
The processing of each data fragment at each step of the job can begin immediately once the processing for the same fragment from the previous step completes. We call this method \emph{Feed Forward} and in this way, we can achieve a significant reduction in the overall processing time for the Job. A user can choose some steps not to be Feed Forward, if he wishes. 
In this case, the processing of all fragments in this step needs to complete in order for the processing to continue to the next step.

As mentioned above, each Job comprises a set of Processing Steps which are developed in the form of containerized Serverless Functions. At any point in time, each step can be only in one of the following states:
\begin{itemize}
    \item \textbf{Pending}, in which the instances of the function have not yet been scheduled.
    \item \textbf{Running}, in which the instances of the function have been scheduled and are executing.
    \item \textbf{Waiting}, in which the function instances have not started as they wait for the previous step to complete. This condition is only encountered in non-Feed Forward Steps.
    \item \textbf{Completed}, which indicates that the processing of this step has been completed and all active instances can be terminated.
\end{itemize}

\section{System Components}
In this section, we present our system architecture that comprises (i) the {\it Batch Driver} and the {\it HCS Scheduler}.

\subsection{Batch Driver}
The Batch Driver is the core component of the Batch Job. It is responsible for both executing and monitoring the Job's progress. A job can be conceptualized as a series of function invocations organized as a DAG, which is expressed by the user in JSON. The JSON file also includes the minimum required resources for each function in the job pipeline. In addition, the JSON file includes a soft deadline which serves as an indicative and feasible timeframe for completing the job. 

The Batch Driver utilizes the Latency Estimator component, which periodically estimates and reports the remaining execution time of the Batch Job. By monitoring the processing rate of each fragment and the number of remaining fragments, the estimator can approximate the remaining time for job completion. Furthermore, the Batch Driver uses a helper component, the Batch Streamer, which is responsible for invoking the functions at each step through HTTP connections. 
The Batch Streamer can maintain communication with different regions simultaneously and dynamically adjust the size of each worker pool. It is also possible to pause and continue processing at any time. The Streamer component initiates its processing once it receives a suitable event from the Batch Driver, indicating that the functions are deployed and ready to handle requests.
In terms of Fault Tolerance, the Batch Driver component uses a consistent journaling mechanism stored in etcd and by checking the results in the storage, (e.g. S3 buckets) it can resume the processing by re-sending only the unprocessed fragments.

\begin{figure}[h]
\label{fig.architecture}
  \centering
  \includegraphics[width=0.8\linewidth, height=3.5cm]{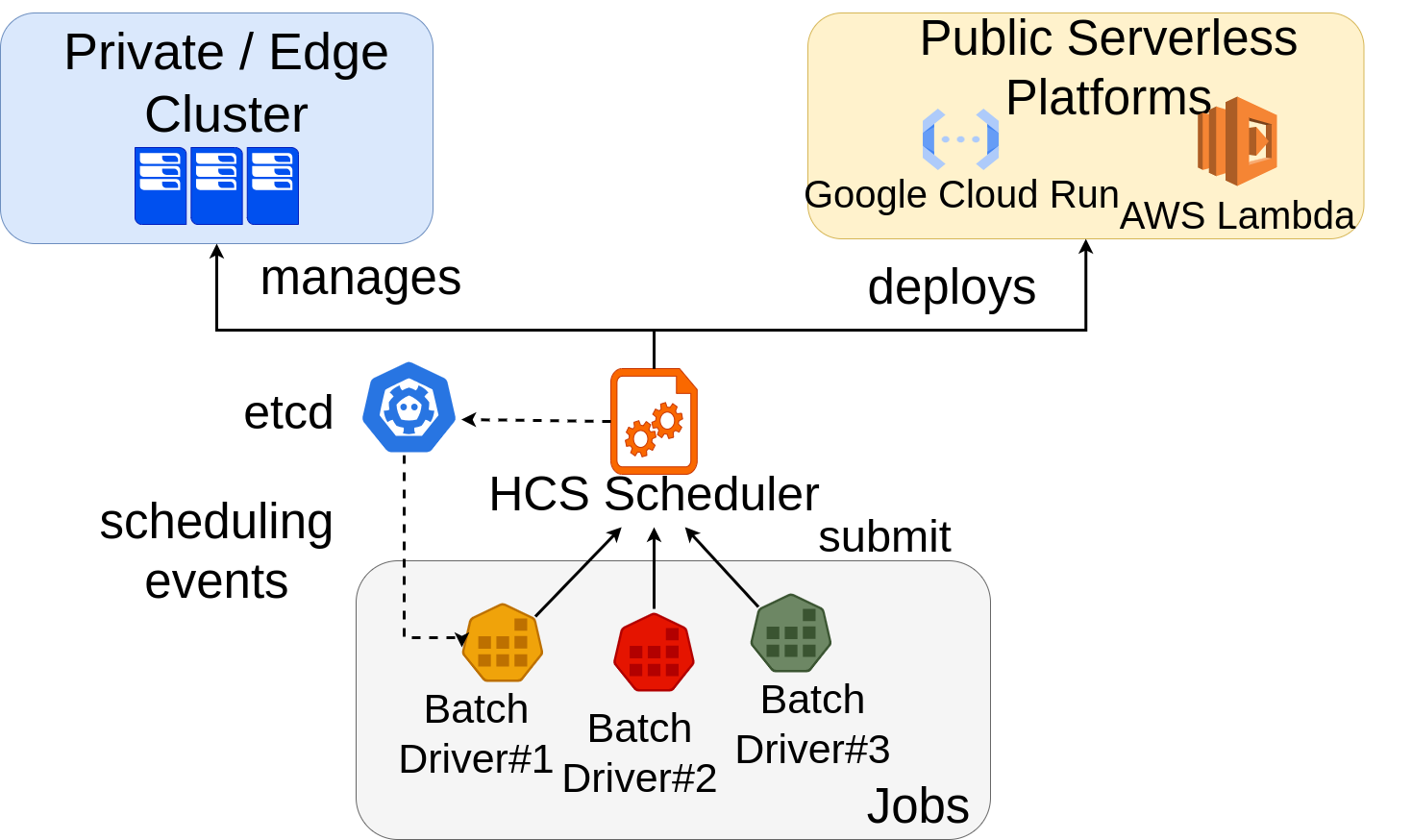}
  \caption{System architecture}
\end{figure}

\subsection{HCS Scheduler}
The Hybrid Cloud Scheduler (HCS) is the key component for our adaptive scheduling approach in the Hybrid Cloud. The HCS Scheduler has full knowledge and control over the private clusters and can deploy functions to the public Cloud. In the public Cloud Serverless container platform, the Scheduler cannot select the number of replicas or the placement of the functions. It needs to be equipped with the appropriate API provided by the public Cloud Serverless container platform to be able to interact with it.
The HCS Scheduler has full knowledge of all the available functions and their parameters in both regions which are provided by the Batch Driver during the initialization of each Job. 
The Streamer component asynchronously receives requests for scheduling batch pipelines. Each request includes the deadline for job completion and the required number of replicas for each function in the Private Edge Cluster to meet the deadline. Requests that have been received in the same round are evaluated together. At the end of each round, a scheduling decision is made and the Jobs are scheduled into the appropriate Serverless execution environment. 
 
If the scheduler chooses to schedule the function in the Private cluster it can allocate the necessary resources and immediately deploy the functions and the appropriate Batch Driver gets notified through etcd, which acts as our event delivery system. The Driver can invoke functions in the pipeline with $m$ concurrent requests, where $m$ represents the number of active replicas. In the Private cluster, $m$ is equal to the number of deployed replicas, as we follow a One Worker Per Request execution model for the functions \cite{amesos}.

In the case that the function is scheduled in the public serverless platform the scheduler emits an event that includes the public URL for that function to the Driver through etcd. The entire amount of resources that are allocated in the public platform are unknown to the scheduler and are automatically adjusted according to the current request rate to the function. In this case, the Driver component can choose an arbitrary number of concurrent requests that can produce the required throughput in order for the pipeline to achieve its required SLOs.

Finally, the scheduler periodically receives heartbeats from the cluster nodes via the container orchestrator. In the event of a node failure in the Edge cluster, the Scheduler will reschedule functions either to other available nodes or to
the Cloud. If the Scheduler itself fails, it can restore the state of both regions by reading the state from the respective container orchestrator; however comprehensive fault-tolerance mechanisms
are planned for future development.

\section{HCS Scheduling policies}

\subsection{Scheduling Methodology}
\label{policies}
\textbf{Jobs:} We consider a set of Batch Jobs where each job needs to execute within a time deadline $T_{deadline}$. Each Job $j$ consists of $N_j$ steps. As we mentioned earlier the jobs data are split into $m$ mini-batches. We consider a \emph{Feed Forward} execution model. In \emph{Feed Forward} a mini-batch can be propagated to the next processing step immediately upon the completion of its current step. Considering these assumptions we can argue that: A processing step can be replicated for increased throughput and all processing steps can be deployed simultaneously.

\textbf{Execution Environment:} The Jobs can be executed either in a private Edge cluster or in a Public Serverless platform, such as AWS Lambda, Google Cloud Functions or Google Cloud Run. The cost of executing a Job as a set of containerized Functions in the private cluster can be considered negligible. On the other hand, deploying Functions to the Public Serverless platform has a variable cost that depends on the total allocated resources for the function execution and the total duration that the function is deployed and executed. To compute the cost, we follow the guidelines of the cloud providers, for example, in AWS Fargate this is defined as follows:

\begin{equation}
\label{TotalCost}
   TotalCost(f_i) = RCost(f_i) \cdot deployedTime_{f_i}
\end{equation}
\vspace{-3mm}
\begin{equation}
\label{Resources Cost}
   RCost(f_i) =  (Memory_{f_i}  \cdot C_m + CPUs_{f_i}  \cdot C_{cpu}) \cdot replicas_{f_i}
\end{equation}

\noindent
where $RCost(f_i)$ is the total cost per second for allocating the $replicas_{f_i}$ with $Memory_{f_i}$ and $CPUs_{f_i}$ resources, where each unit costs $C_m$ and $C_{cpu}$ accordingly and $deployedTime_{f_i}$ is the total time that the resources are allocated. 

\subsection{Scheduling Goal}

The goal of our scheduling approach is two-fold: (i) reduce the total execution cost for a set of Batch Jobs, without violating their deadlines by efficiently utilizing the private Edge cluster, and (ii) exploit different function placement policies to increase utilization in the Edge cluster. The Edge cluster has limited resources, thus it may not be possible for all the Jobs to fit in parallel in the Cluster. To achieve their respective deadlines, some of the Jobs or part of the Jobs should be deployed to the public Serverless platform. 

The scheduler operates in rounds, it collects scheduling requests and then executes the scheduling policy. Then the scheduler, for each Job affected, informs the appropriate Batch Driver with the new endpoint of each function. For functions scheduled in the public platform, the scheduler returns the cloud provider's published endpoint. Suppose the scaling decision indicates a scale-down for some functions, in that case, the scheduler upon emitting the event waits for a \emph{graceful shutdown period} before stopping the function instances. 

\subsection{Scheduling Policy}

In order to reduce the overall execution cost of a set of Batch Jobs without violating their deadlines, we have implemented an efficient scheduling policy. 
For our scheduling policy, we propose the \emph{Cheapest First}. With this policy, a job's function will be deployed at the edge if all of its replicas fit. If there is insufficient capacity, the scheduler checks for cheaper alternative functions and reschedules them in the cloud, prioritizing cost efficiency. The benefit of this technique is that it lowers the price we have to pay in the cloud by keeping on the edge the most expensive functions. 
To minimize unnecessary function movements and maintain pipeline stability, once a job is scheduled to the cloud, it will not be rescheduled back to the edge even if it can fit later on. This approach reduces the frequent movement of functions with lower execution costs because if the edge fills again those same functions will have to be moved  back to the cloud. By avoiding constant transitions between the cloud and edge, the overall performance and stability of the pipeline are improved. 
The scheduler informs the applicable Drivers when no function transfer is required and proceeds with the schedule. However, if a function transfer is necessary, the scheduler emits an event to the corresponding Drivers. The Drivers have an eviction deadline to gracefully shut down pending requests. After the deadline expires, the scheduler terminates function instances in the edge cluster, deploys the new functions in their place, and reschedules any terminated or non-fitting functions in the public Serverless platform. Once the schedule completes, an event is sent to the Drivers to inform them about their functions.

\subsection{Function placement policies in the Private cluster}
As mentioned above, one of the scheduling goals of our approach is to improve the utilization of our private cluster. To do that, we propose different function placement policies at the Edge. Each Function is shipped as a Docker container with a predefined resource demand ({\it e.g,} 0.5 CPU and 128Mb of memory). This demand is the minimum required for the Function to operate properly and cannot be reduced. Due to varying resource requirements, some cluster nodes may have limited availability. The task of selecting appropriate nodes for deployment is similar to the NP-Hard Bin-Covering problem, for which approximate solutions exist \cite{Boyar2020} \cite{mig} but may involve impractical shuffling or require advice for achieving good competitive ratios.

For our scheduler, we propose four simple greedy Function placement policies used in bin-packing. The first approach utilizes the \emph{First-Fit} (FF) technique of bin packing. In this approach when new jobs arrive for scheduling we pick the first available node and start placing the containerized functions until it reaches its maximum capacity. The second placement policy is the \emph{Best-Fit} (BF) technique of bin packing. As new jobs arrive for scheduling, the scheduler selects the node with the fewest available resources. The third placement policy is the \emph{Round Robin} (RR) approach. In this approach when a new job arrives for scheduling, each function is deployed in a round-robin fashion across all available nodes in the cluster. Finally, the \emph{Worst-Fit} (WF) works similarly to the BF approach but when a new job arrives for scheduling, WF picks the node with the most available resources where it fits. 
\label{policy}

\section{Implementation}

We have implemented our system, we created the Scheduler and Batch-Driver template that can be used to create batch processing pipelines by defining a DAG. As the underlying container orchestrator for the two clusters, we used Apache Mesos with Mesosphere Marathon. The Scheduler overrides the default scheduling policy of Marathon for the function placement in the Edge Cluster. For function invocations, a custom reverse proxy gateway has been used. The reverse proxy is not needed in the case we invoke functions in a public Cloud such as Google Cloud Run. As object storage, we used MinIO, which is an open-source S3-compatible object storage. Etcd key-value store used by the scheduler for informing the Batch-Drivers for scheduling decisions via its event delivery system. Finally, Prometheus is used for monitoring the progress of the Batch Jobs that are running in the system at each time. The Batch Drivers push their progress via the Prometheus Push Gateway and the Prometheus Server periodically collects them. Also, the Scheduler reports to Prometheus the state and the utilization of each region, as well as the total cost of the job execution.

\section{Experimental Evaluation}

\subsection{Experimental Setup}

\textbf{Hardware}.
To conduct our experiments we split our local cluster into two regions, one region for the Edge and another for simulating the Cloud. In both regions, each node has an Intel i7-7700 3.6GHz processor with 4 physical cores and 8 threads and 16GB of RAM. All of the nodes and VMS are interconnected with 1GBps Ethernet and run on Ubuntu 20.04 LTS.
The Edge cluster comprises 6 virtual machines running with 3 vCPUs and 10GB of RAM each. At each virtual machine, we set an execution cap of 80\% of the CPU usage to differentiate them from the machines that are in the Cloud region.

\begin{figure*}[h]
\centering
  \includegraphics[width=0.8\textwidth,height=3cm]{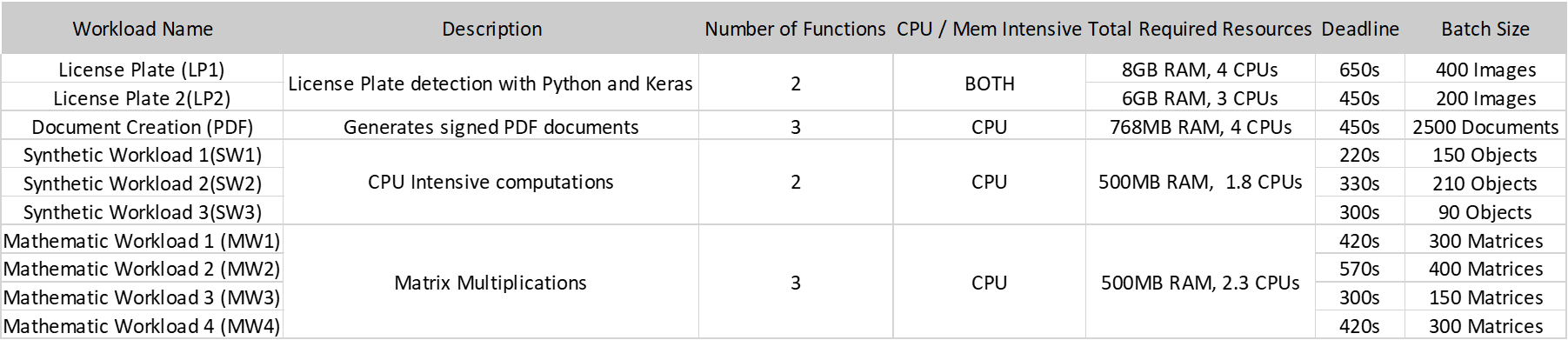}
  \caption{Workload Description}
  \label{lable.table}
\end{figure*}

\textbf{Functions}. We used OpenFaas Python templates to create our functions. We organized our functions into four different Pipelines as shown in Table \ref{lable.table}. 
The \emph{License Plate} detection pipeline uses YOLOv5s \cite{yolo} for car detection and WPOD-NET \cite{wpod} for license plate detection in street images. It stores the car bounding box positions and performs character segmentation using OpenCV to extract and save license plate characters in a text file, which is then stored in MinIO. The \emph{Document Creation} pipeline creates signed PDF documents. It consists of three different functions: SHA256, which generates the digital signature of the text, Qrcode which generates a QRCode image that contains the signed and Pdf which combines the signed text and the generated Qrcode into a single PDF file. The two other types of workloads, \emph{Synthetic Workload} and \emph{Mathematic Workload} are synthetic. They consist of two and three functions respectively and they perform CPU-intensive operations in a number of objects stored in MinIO buckets. The Synthetic Workload performs AES Encryption and Decryption cycles on each object and the Mathematic Workload performs a series of matrix multiplications for each matrix polled from MinIO.  

\begin{figure*}[ht]
\minipage{0.32\textwidth}
  \includegraphics[width=0.8\linewidth]{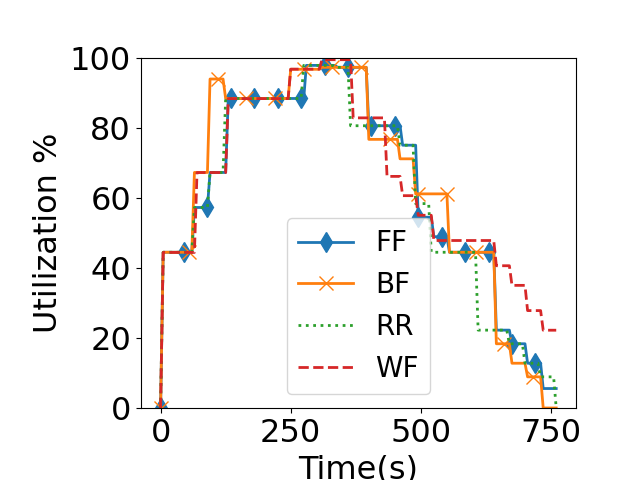}
  \centering
   \caption{Edge cluster utilization}
\endminipage\hfill
\minipage{0.32\textwidth}%
 \includegraphics[width=0.8\linewidth]{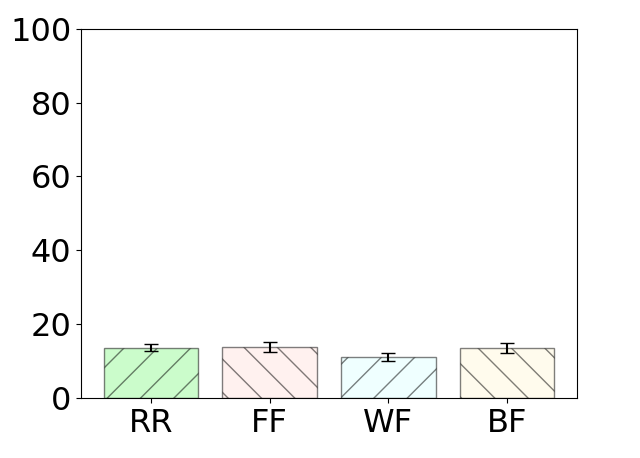}
   \centering
   \caption{Cost percentage compared to baseline}
\endminipage
\minipage{0.32\textwidth}
   \includegraphics[width=0.9\linewidth]{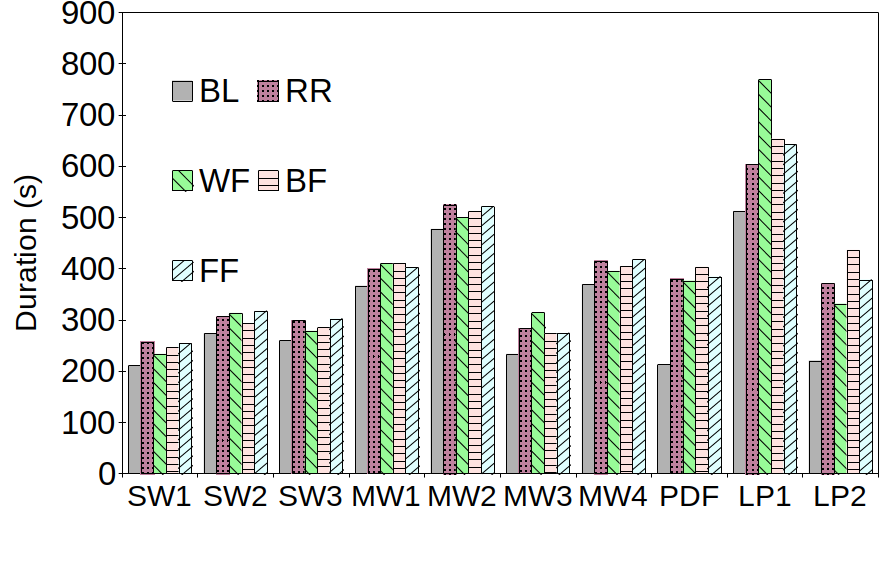}
     \centering
   \caption{Workload execution duration}
\endminipage\hfill
        \label{fig:result-util}
\end{figure*}

\subsection{Evaluation Method}
We designed and conducted experiments in order to evaluate our scheduling policy in terms of cost reduction and execution performance compared to  deploying the workloads only to the cloud. Furthermore, our experiments are designed to compare the different placement policies that are proposed in section \ref{policy} in terms of edge utilization.

The workloads arrive following a Poisson distribution. We use a known seed so that the experiments are repeatable and have similar behaviour across the different policies that are tested. The Docker images of the functions are pre-fetched to all the nodes, both in the Edge and Cloud cluster in order to reduce the unpredictable cold start duration due to the limited network bandwidth available. 

For our baseline approach, we use a Cloud Only Scheduling Policy. For our proposed Scheduling Policy we repeated the same experiment for all the Placement Policies, three times and we calculated the average results. This was done in order to mitigate the small differences in the execution times that might occur between each scenario due to the different waiting duration of each workload until the scheduling round finish and the Scheduler triggers a scheduling action. In our experiments, the scheduling round is set to 30s and the eviction deadline is also set to 30s. For calculating the execution cost of each Function we use \ref{TotalCost} and \ref{Resources Cost} with $ C_{cpu}=1000 $ and $ C_{m}=0.1$, as described in Section \ref{policies}. 

\subsection{Evaluation Results}

We evaluate our approaches using the following metrics:

\textbf{Edge Cluster Utilization:} In Figure \ref{fig:result-util} (A) we report the edge cluster utilization. In this figure, we plot only the CPU utilization, this is because the overall CPU demand is higher than memory and the Edge cluster runs out of CPU resources first.
The scheduling policy achieves high utilization, above 90\%, of the Edge resources across all the various placement policies.

\textbf{Execution Cost compared to Baseline:} In Figure \ref{fig:result-util} (B) we illustrate the total cost of all the workloads compared to the baseline approach, which schedules them all to the Cloud. Our approach achieves high cost reduction, as much as 87\% cheaper compared to the baseline cost.

\textbf{Workload Duration:} In Figure \ref{fig:result-util} (C) we depict the execution times of all the workloads. As anticipated, the baseline execution duration time is lower than the duration of the approach that uses the edge. In our experiment, this occurs because the Edge cluster virtual machines have lower performance than the machines in the Cloud Region. Also in a real-world scenario, typically the machines that run in public data centers have hardware of higher performance compared to private clusters. Despite this performance difference, the edge cluster performance was still good enough to meet almost all execution deadlines which were set approximately 60s higher than the expected execution time. In general, all the approaches had similar deadline misses with our approach missing the deadline of SW2 in all the placement policies by 30s and by 110s only for the LP1 in the WF policy.

\section{Related Work}
A lot of work have been conducted in the area of scheduling workloads in Serverless environments \cite{energy} \cite{tomaras2}.
Hourglass \cite{hourglass} is a system that utilizes transient resources like AWS EC2 spot instances for cost-effective, time-constrained graph processing. It employs a slack-aware provisioning strategy and a fast reload mechanism to handle spot instance evictions efficiently.
Van den Bossche et al. in \cite{Vandoro} combine private infrastructure with resources from public cloud providers to execute cost-efficient scheduling for time-sensitive workloads. To achieve this in their work, their scheduler takes into consideration data locality, data constraints, runtime estimate inaccuracies and data transfer cost. Zhang et al. in \cite{fasterAndCheaper} try to reduce the execution cost for the Serverless provider by hosting Serverless Functions on harvested virtual machines which can grow and shrink to harvest all the available resources in their host servers but may be evicted to make room for more expensive VMs. Mashup \cite{Mashup} combines AWS Lambda Functions and EC2 to create a hybrid HPC workflow execution system. It selects the platform with the lowest latency for each task based on performance profiling and utilizes serverless computing to improve execution time and cost efficiency in a hybrid environment. Kaffes et al. in Hermod \cite{kafes} explores scheduling techniques for serverless systems. They propose a cost, load and locality-aware scheduler in order to reduce the number of cold starts compared to pure load-based policies and maintain high performance. In a different approach, FaaSRank \cite{faasRank} used Reinforcement Learning (RL) to automatically learn the scheduling policies through experience in clusters running serverless functions. Giuseppe De Palma et al. \cite{italian} propose a declarative language to define the policies that guide the scheduling of function execution in IoT, Edge and Cloud Computing environments.

\section{Conclusion}
In this paper, we have presented a novel framework for executing serverless Batch jobs in Hybrid Clouds consisting of private and public resources. We have implemented a two-level Hybrid scheduler with four placement function policies for the private cluster. Our evaluation shows up to 87\% cost reduction to using public serverless platforms only.

\vspace{-3mm}
\section*{Acknowledgement}
The research work was supported by the Hellenic Foundation for Research and Innovation (HFRI) under the 3rd Call for HFRI PhD Fellowships (Fellowship Number: 6812), by the European Union through the EU ICT-48 2020 project TAILOR (No. 952215), the H2020 AutoFair project (No. 101070568) and the Horizon Europe CoDiet project (No. 101084642).

\bibliographystyle{IEEEtran}
\bibliography{refs}

\end{document}